\documentclass[preprint,showpacs,preprintnumbers,amsmath,amssymb]{revtex4}
\usepackage{amsmath,amssymb,graphicx} 
\usepackage{epsf}
\usepackage{epstopdf}
\usepackage {amssymb}
\newcommand{\nc}{\newcommand}
\nc{\ba}{\begin{eqnarray}}
\nc{\ea}{\end{eqnarray}}
\nc{\rb}{\bar{\rho}}
\nc{\p}{\phi}
\nc{\la}{\lambda}
\nc{\al}{\alpha^{\prime}}
\nc{\de}{\delta_{H}}
\nc{\be}{\begin{equation}}
\nc{\ee}{\end{equation}}
\nc{\D}{\overline{\mbox{D3}}}
\newcommand{\e}{\epsilon}
\newcommand{\dd}{\Delta}
\newcommand{\oo}{\Omega}
\def\ap{{\alpha^{\prime}}}

\newcommand{\STr}{{\rm STr}}
\newcommand\hi{{\rm i}}

\input{epsf.sty}

\begin{document}


\title{Dielectric (p,q) Strings in a Throat}

\author{Hassan Firouzjahi}
\email{firouz@physics.mcgill.ca}
\affiliation{Physics Department, McGill University, 3600 University Street, Montreal, Canada, H3A 2T8 }

\begin{abstract}
We calculate the (p,q) string spectrum in a warped deformed conifold using the dielectric brane
method. The spectrum is shown to have the same functional form as in the 
dual picture of a wrapped D3-brane with electric and magnetic fluxes on its world volume. The agreement is exact in the limit where q is large. We also calculate the dielectric spectrum in the S-dual picture. The spectrum in the S-dual picture has the same form as in the original picture but
it is not exactly S-dual invariant due to an interchange of Casimirs of the
non-Abelian gauge symmetries.
We argue that in order to restore S-duality invariance the non-Abelian brane action should be 
refined, probably by a better prescription for the non-Abelian trace operation.

\vspace{0.3cm}

\end{abstract}

\maketitle

\section{Introduction}

Recent observations strongly support inflation as the origin of the universe and the structure formation
\cite{Spergel:2006hy}. Despite its impressive success in explaining a host of observational data such as
the CMB power spectrum and in solving conceptual puzzles such as the flatness and the horizon problems, the inflationary scenario is still at the phenomenological level. The origin of the inflation and the nature of the inflaton field is not known from a fundamental theory point of view. String theory, on the other hand, is a consistent theory of quantum gravity which is yet to be tested. 
The energy scale of inflation was probably so high that quantum gravity effects were important, either directly or indirectly.
It seems reasonable to expect that string theory, assuming it to be relevant
to our universe, will yield many insights into the nature of inflation. If this notion turns out to be true, it would be a unique chance
to test the relevance of string theory to our universe.

String theory is a higher dimensional theory. After compactification to four dimensions, 
many scalar fields
or moduli will show up in low energy theory. This has both positive and negative sides.
The positive side is that some of these moduli may play the role of the inflation field. In this view, inflation is a natural feature of string theory compactifications. The negative side is that usually many of these moduli will roll quickly. This may violate the slow roll conditions, required for slow roll based inflation.
Furthermore, if they are not stabilized and roll quickly they may modify the expansion history
of the universe. This can destroy the success of the late time big bang cosmology such as the predictions of the big bang nucleosynthesis for the abundance of the light elements. In this regard the moduli stabilization is a key issue in string cosmology.

There has been interesting progress in moduli stabilization  \cite{Dasgupta:1999ss},  \cite{Giddings:2001yu},  \cite{Kachru:2003aw}.
By turning on fluxes it was shown in \cite{Giddings:2001yu} that, in principle, all complex structures can be stabilized. Some non-perturbative mechanisms  are used in \cite{Kachru:2003aw} to stabilize the Kahler modulus, the field corresponding to the overall size of the Calabi-Yau (CY) compactification.
In the constructions of \cite{Giddings:2001yu} and  \cite{Kachru:2003aw} there are localized regions inside CY where the geometry of the space-time is warped similar to Randall-Sundrum scenario \cite{Randall:1999ee}. The particular model studied is the warped deformed conifold 
solution of Klebanov-Strassler (KS) \cite{Klebanov:2000hb} glued smoothly to the bulk of the CY manifold.
In the language of  Randall-Sundrum, the IR region is the bottom of the conifold whereas the
 UV region is the place where
the conifold is smoothly glued to the bulk. The hierarchy of scales between IR and UV is exponentially
sensitive to the ratio of fluxes. This way a huge hierarchy of scales can be created with minor tuning of
fluxes.

An interesting mechanism to obtain inflation in this set up is to use the brane-antibrane inflationary 
scenario \cite{dvali-tye}, \cite{collection} inside the KS throat \cite{Kachru:2003sx}, \cite{Firouzjahi:2003zy}.
 The $\D$-brane is sitting at the IR region of 
the throat while the D3-brane is moving toward it inside the throat. Due to warp factor, the potential
between D3-$\D$ becomes flat enough that the slow roll conditions can easily be satisfied. 
However, the extra contribution from the Kahler modulus of the CY compactification to the inflationary potential conflicts with the slow conditions. It was argued in  \cite{Kachru:2003sx}
that a fine-tuning of order one percent is required to keep the inflationary potential
flat. Inflation ends when brane and anitbrane annihilate each other and this annihilation releases
the energy to reheat the universe. After the branes annihilation, fundamental strings( F-strings) and
D1-brane(D-strings) are copiously produced. These F and D strings have cosmological size
and appear as cosmic strings \cite {Sarangi:2002yt}, \cite{Jones:2003da}, \cite{Copeland:2003bj}.
The range of their tension is compatible with the recent observations \cite{Firouzjahi:2005dh}, 
\cite{Shandera:2006ax}, \cite{Wyman:2005tu}. The cosmic strings tension in these models is
high enough to be detected in near future gravity wave search \cite{Seljak:2006hi}. Detecting cosmic
strings would go a long way in support of brane inflation and string theory. 
 
Since the branes annihilation takes place in the IR region of the KS throat, the cosmic strings
are localized in this region, near the tip of the KS throat. The spectrum of cosmic strings is a rich combination of bound states of F and D-strings. It is an interesting question to
understand the spectrum of $(p,q)$ strings, the bound states of $p$ F and 
$q$ D-strings on top of each other at the tip of the throat. This question was studied recently 
in \cite{Firouzjahi:2006vp}. In this method, $(p,q)$ strings were viewed as a D3-brane with $p$
units of electric flux and $q$ units of magnetic flux in its world volume, wrapping a two-sphere
inside the three-sphere at the tip of the KS throat. The $(p,q)$ spectrum was shown to be in agreement
with the flat space formula. Furthermore, in the case $q=0$ it reproduces the result of  \cite{Herzog:2001fq}
for tension spectrum of $p$ F-stings in $SU(M)$ gauge theory.

One may try to calculate the $(p,q)$ spectrum directly, using the dielectric brane method   
\cite{Myers:1999ps}. In this picture, $q$ D-strings on top of each other blow up into a fuzzy
D3-brane, while the fundamental strings are viewed as $p$ units of electric flux on D-strings world volume. The extra two coordinates on the world volume of the fuzzy D3-brane corresponds
to the positions of $q$ D-strings in the orthogonal directions. These orthogonal dimensions 
span a two-sphere
inside the three-sphere at the tip of the throat. In a sense, the dielectric brane method of calculating
the $(p,q)$ spectrum is the reverse direction of calculating the $(p,q)$ spectrum using the 
wrapped D3-brane method in \cite{Firouzjahi:2006vp}. In general, one expect that for large $q$ these two methods give the same result for the
spectrum  \cite{Myers:1999ps}, \cite{Myers:2003bw}.

The paper is organized as follows. In next section we briefly review the KS throat. In section 3 the
$(p,q)$ spectrum is calculated using the dielectric brane method. In section 4 this calculation 
is carried on for the S-dual background of KS throat. The Conclusions are given in section 5.

\section{The warped deformed conifold}

In this section we briefly review the warped deformed conifold.  A cone is defined by the following equation in ${\cal{\bf{C^4}}}$ \cite{Candelas:1989js}
\ba
\label{6cone}
\sum_{i=1}^{4} w_i^2=0 \, .
\ea
This describes a smooth surface apart from the point $w_i=0$. 
The base of the cone is 
given by the intersection of  Eq.(\ref{6cone}) with a sphere of radius $r$ in $R^8$, $$\sum_i |w_i|^2 = r^2$$

We are interested in Ricci-flat metrics on the cone which in turn imply that the base of the conifold is a 
Sasaki-Einstein manifold.  The simplest five dimensional Sasaki-Einstein manifold for $N=1$ supersymmetry is $T^{1,1}$ and it is the only manifold for which the deformation is explicitly known \cite{Klebanov:2000hb}.

The metric on the conifold with base $T^{1,1}$ is 
\ba
\label{6t11}
ds_{6}^{2}&=& dr^{2} + r^{2} ds_{T^{1,1}}^2 \,   \\
ds_{T^{1,1}}^2 &=&\frac{1}{9}(\,d\psi +\sum_{i=1}^{2} 
\cos \theta_i\, d \phi_i\, )^2 + \frac{1}{6}\sum_{i=1}^{2}
 (\, d\theta_i^2 +  \sin^2 \theta_i\, d\phi_i^2\,)\, . \nonumber  
\ea
It can be shown that $T^{1,1}$ has topology of $S^2 \times S^3$ with $S^2$ fibered over $S^3$. 
If $\varphi_1$ and $\varphi_2$ are the two Euler angles of the two $S^{3}$'s, 
respectively, then their difference corresponds to $U(1)$ while $\psi=  \varphi_1+ \varphi_2$. 
Since $2 \pi \ge \varphi_i \ge 0$, the range of $\psi$ is $[0, 4 \pi]$.

The Klebanov-Strassler throat that we are interested in is actually a warped deformed conifold. This warped deformed conifold emerges in the presence of fluxes.
The R-R flux $F_3$ wraps the $S^{3}$ at the bottom of the deformed conifold, while NS-NS flux $H_3$ wraps the dual 3-cycle $B$ that generates the warped throat  
\ba
\label{KMNhA}
\frac{1}{ 4\pi^{2}\ap}\int_B H_3 &=& -K, \quad \quad 
\frac{1}{ 4\pi^{2} \ap}\int_{S^{3}} F_3 = M \, . \nonumber \\ 
\ea

The metric of the deformed conifold is studied in 
\cite{Minasian:1999tt,Ohta:1999we,Herzog:2001xk}.
At the tip of the deformed conifold one $S^2$ shrinks to zero size and the internal geometry reduces to
a round $S^3$. At the tip the metric is given by 
\ba
\label{metric0}
ds^2 \sim  h^2\, \eta_{\mu \nu} dx^{\mu} dx^{\nu}+
 b\, g_s M \alpha'(d\psi^2 +\sin^2 \psi\, d \oo_2 ^2)
\ea
where $h$ is the warp factor at the bottom of the throat
\ba
h= \e^{2/3}2^{-1/6} a_0^{-1/4} (g_s M \alpha')^{-1/2}
\ea
 where $a_0 \sim .72$ and $b= 2^{2/3} 3^{-1/3} I(0)^{1/2}\sim 0.93$ are numerical constants in the KS solution and $\e^{1/3}$ is the deformation radius. 
Here $\psi$ is the usual azimuthal coordinate in a $S^3$ ranging from $0$ to $\pi$.

There are $M$ units of RR 3-forms $F_3$
on the non-vanishing $S^3$ cycle at the tip of the throat. Its associated two form is given by
\ba
\label{C2}
C^{(2)}=  M \alpha' \,  \left(\psi-\frac{\sin (2\psi)}{2} \right) \sin \theta\, d \theta \, d\phi \, .
\ea
In KS solution $C_0=0$ and at the bottom of the throat $B_{ab}=0$.

\section{The Dielectric (p,q) Strings}

One may consider to obtain the bound states of $(p,q)$ strings directly \cite{Thomas:2006ud},
using the non-commutative dielectric brane method prescribed by Myers \cite{Myers:1999ps}. 
Upon expansion, the results of \cite{Thomas:2006ud} were in agreements with the results of \cite{Firouzjahi:2006vp} up to order $1/M^2$. It is an interesting exercise to see wether there is also an exact agreement between the
non-commutative dielectric brane method and the dual method of wrapped D3-brane
used in \cite{Firouzjahi:2006vp}.

When N p-branes are located on top of each other, the ground state of open strings attached between them becomes massless and the $U(1)^N$ symmetry associated with N individual branes is enhanced to $U(N)$. The gauge field vector $A_a$ becomes non-Abelian and
\ba
A_a= A^{(n)}_a T_n  \quad \quad , \quad \quad F_{ab}= \partial_a A_b - \partial_b A_a + i [ A_a, A_b] \, ,
\ea
where $T_n$ are $N^2$ Hermitian generators with $Tr (T_n T_m) = N \delta_{n m}$. The orthogonal displacements of branes, $\Phi^i$, are now matrix valued and transform in the adjoint of U(N) with
\ba
D_a \Phi^i = \partial_a \Phi^i + i [A_a, \Phi^i] \, .
\ea

The non-Abelian action of $N$ coincident $p$-branes is given by $S=S_{DBI} +S_{CS}$, where
 \cite{Myers:1999ps}
\ba
\label{SDBI}
S_{DBI}=-\mu_p \int d^{p+1}\sigma\,\STr \left(e^{-\phi}\, \sqrt{\det(Q^i{}_j)}  \sqrt{-\det\left(
P\left[E_{ab}+E_{ai}(Q^{-1}-\delta)^{ij}E_{jb}\right]+
\lambda\,F_{ab}\right)} \right) 
\ea
and
\ba
\label{SCS}
S_{CS}=\mu_p\int \STr\left(P\left[e^{i\lambda\,\hi_\Phi \hi_\Phi}
( \sum C^{(n)}\,e^B)\right] e^{\lambda\,F}\right)\ .
\ea
In these expressions, $\lambda= 2\pi \ap$, $\mu_p$ is the p-brane charge,
 $E_{M N}=g_{M N}+B_{M N}$, $Q^i{}_j\equiv\delta^i{}_j+i\lambda\,[\Phi^i,\Phi^k]\,E_{kj}$ and $\Phi^i$ is the branes location along the orthogonal direction $x^i$. In the conventions of
\cite{Myers:1999ps}, $x^i= \lambda \Phi^i$, so that $\Phi^i$ has dimensions of mass.
The indices $a, b,...$ are along the brane 
world-volume directions, while $i, j, ...$ represents the directions orthogonal to the branes. The operation 
STr corresponds to the trace of symmetrised pairing of the non-Abelian fields $F_{ab}$ and $\Phi^i $ 
\cite{Tseytlin:1997cs}.
All induced quantities are the pull-backs of space-time tensors on the brane.

In $S_{CS}$, the operator $\hi_ \Phi$ denotes the interior product acting on an n-form 
$C^{(n)}= \frac{1}{n!} C^{(n)}_{M_1...M_n} dx^{M_1}...dx^{M_n} $ and
\ba
\hi_\Phi\hi_\Phi C^{(n)}={1\over2(n-2)!}\,[\Phi^i,\Phi^j]\,
C^{(n)}_{ji M_3\cdots M_n}dx^{M_3}\cdots dx^{M_n} \, .
\ea

To calculate the $(p,q)$ spectrum, one can turn on $p$ units of $U(1)$ electric field on the world volume of $q$ coincident D1-branes. We take the D1-branes to be extended in the $(t, x^\mu)$ direction and the only relevant component of $F_{ab}$ is $F_{0\mu}$.
Furthermore, at the tip of KS throat $B_{M N}=0$ and $E_{M N}= g_{M N}$. The only non-zero term in 
$S_{CS}$ is the term coming from the operation of $\hi_ \Phi$ on $C^{(2)}$.

With these assumptions the action of q D1-branes with electric fields on their world volumes is
\ba
S=\int d\,t \, d\,x^{\mu} \,\left [-\frac{\mu_1}{g_s} \sqrt{h^2-\lambda^2 F_{0\mu}^2 } \,\,\, \STr \sqrt{\det(Q^i_j)} +  
\frac{i \lambda}{2} F_{0 \mu}\, \STr \left( [\Phi^i, \Phi^j] C^{(2)}_{ji}  \right)\, .
\right]
\ea
To simplify the algebra it is useful to write the above action in the form
\ba
\label{def1}
S= \int  d\,t \, d\,x^{\mu} \, \left[\, -\dd \sqrt{h^{4} - \lambda^2F_{\mu 0}^2} + \oo F_{0 \mu} \right] \, ,
\ea
with
\ba
\label{def2}
\Delta\equiv -\frac{\mu_1}{g_s} \, \STr \sqrt{\det(Q^i_j)} \quad, \quad 
\Omega\equiv \frac{i \lambda}{2}\, \STr \left( [\Phi^i, \Phi^j] C^{(2)}_{ji}   \right)
\ea
The conjugate momentum associate with the electric field is
\ba
D=\frac{\delta {\cal{L}}}{\delta{F_{0 \mu}}} =  \frac{\dd \lambda^2 F_{0 \mu}}{\sqrt{h^{4} -\lambda^2 F_{\mu 0}^2}} + \oo \, .
\ea
Using this expression to eliminate $F_{0 \mu }$, the Hamiltonian is \cite{Firouzjahi:2006vp}
\ba
\label{Ham}
{\cal{H}} &=& D F_{\mu0} - {\cal{L}} \nonumber\\
& =& \frac{h^{2}}{\lambda}\sqrt{\Delta^2\lambda^2 + (D-\oo)^2}
\ea
Our goal is to calculate $\dd$ and $\oo$ for the model at hand and plug them in the above expression 
for the Hamiltonian. Furthermore, since the F-strings in $(p,q)$ strings are associated with the electric fields this means $D=p$.

Motivated by the dual wrapped D3-brane picture, we expect that $q$ D1-branes blow up into a 
fuzzy D3-brane. This fuzzy D3-brane wraps a two sphere 
inside the $S^3$ at the tip of the deformed conifold. 
The transverse coordinates $x^i$ span the surface of this two
sphere, so $\Phi^i$ belongs to a $q$-dimensional representation of the $SU(2)$ algebra. We take the following ansatz for the transverse coordinates \cite{Constable:1999ac}
\ba
\label{Phi}
\Phi^i= \hat R \, \alpha^i
\ea
where the constant $\hat R$ with dimension mass will be determined later and $\alpha^i$ 
are the $q\times q$ dimensional matrix representation of $SU(2)$ algebra
\ba
[\alpha^i, \alpha^j]= 2 i \epsilon_{i j k} \alpha^k
\ea

 It is important to realize
that the metric on the space spanned by the transverse coordinates $x^i$ is flat and 
$g_{kj}= \delta_{kj}$. The radius of this $S^2$ is basically the same as the radius of $S^2$ given in Eq. (\ref{metric0}), i.e. 
\ba
\label{R}
R^2=\sum_{i=1}^3 (x^i)^2= b \, g_s  M\,  \ap \sin^2 \psi
\ea
On the other hand
\ba
R^2=\frac{\lambda^2}{q} \sum_{i=1}^{3} Tr[(\Phi^i)^2]= \lambda^2 C {\hat R}^2
\ea
which can be used to fix the constant $\hat R$
\ba
\hat R= \frac{R}{\lambda \sqrt C} \, .
\ea
Here $C$ is the Casimir of the representation. For q-dimensional irreducible representation of $SU(2)$
which will be used in our calculation, $C=C_q= q^2-1$.

With these assumptions, we obtain
\ba
\label{Q}
Q^i_j= \delta^i_j - 2 \epsilon_{ijk} \lambda \hat R \,\Phi^k
\ea
and
\ba
\label{detQ}
\det(Q^i_j)= 1+ 4\, \lambda^2\, {\hat R}^2\, \sum_{i=1}^3 (\Phi^i)^2 + ...
\ea
where the extra terms in Eq.(\ref{detQ}) are terms which are not symmetric under exchange of
$\Phi^i$ and do not contribute in $\STr$ operation. One finds
\ba
\label{detQ2}
\Delta= \frac{\mu_1}{g_s}\, \STr \sqrt{\det(Q^i_j)}&=& \frac{\mu_1}{g_s}\, \left(1+ 4\, \lambda^2\, {\hat R}^2\, \sum_{i=1}^3 (\Phi^i)^2 \right)^{1/2} \nonumber\\
&=& \frac{q \,  }{\lambda g_s} \, \left(1+ (\frac{b g_s M}{\pi\, \sqrt{C_q}})^2 \, \sin^4 \psi \right )^{1/2}
\ea
where in the last line the relation $\mu=\lambda^{-1}$ is used and $R$ is given by Eq.(\ref{R}).

To calculate $\Omega$, one needs to transform $C^{(2)}$ in Eq.(\ref{C2}) from
$C^{(2)}_{\theta \phi}$ components  into the Cartesian coordinates $C^{(2)}_{ij}$. 
For example, to calculate $C^{(2)}_{12}$, we have
\ba
C_{12}&=& \left( \frac{\partial\, \theta}{\partial x^1}\,  \frac{\partial\, \phi}{\partial x^2} + 
\frac{\partial\, \phi}{\partial x^1}\,  \frac{\partial\, \theta}{\partial x^2} \right) \, C_{\theta \phi}
\nonumber\\
&=& \frac{x_3\, C_{\theta \phi}}{ R^2 \sqrt{ (x_1^2+x_2^2) }} 
\ea

Performing this coordinate
transformation one obtains
\ba
\left(  C^{(2)}_{12} , C^{(2)}_{23} , C^{(2)}_{31}     \right)= 
 \frac{\lambda }{R^3} \, M \ap (\psi - \frac{1}{2} \sin 2\psi)\, (\Phi_3, \Phi_1, \Phi_2)
\ea 
and
\ba
\epsilon_{ijk} \Phi^k C^{(2)}_{ij}= \frac{2}{\lambda R} (\psi - \frac{1}{2} \sin 2\psi)\, I_q
\ea
where $I_q$ is the q-dimensioanl unit matrix. This gives
\ba
\oo &=& \frac{i \lambda}{2}\, \STr \left( [\Phi^i, \Phi^j] C^{(2)}_{ji}   \right)\nonumber\\
&=&\frac{ q M  }{\pi \sqrt C_q} (\psi - \frac{1}{2} \sin 2\psi)
\ea

Having calculated $\dd$ and $\oo$ and setting $D=p$, we can use Eq.(\ref{Ham}) to calculate the Hamiltonian of
the system
\ba
\label{Ham2}
{\cal {H}}= \frac{h^2}{\lambda} \left({ \frac{q^2}{g_s^2} + \frac{1}{1-1/q^2} \frac{b^2M^2}{\pi^2} \sin^4 \psi
+\left[p- \frac{1}{\sqrt{1-1/q^2}}\frac{M}{\pi}
\left(\psi -\frac{\sin2\psi}{2}\right)  \right]^2} \right)^{1/2}
\ea

The stable solutions, obtained by minimizing the Hamiltonian, are 
\ba
\label{min2}
\left(\psi +\frac{b^2-1}{2} \sin2\psi\right) = \sqrt{1-1/q^2} \, \, \frac{p\, \pi}{M} 
\ea
and the energy of the $(p,q)$ system, $T_{(p,q)}$ , is
\ba
T_{(p,q)}={\cal{H}}_{min} = \frac{h^2}{\lambda} 
\left( {  \frac{q^2}{g_s^2} + \frac{1}{1-1/q^2} \,  \frac{b^2 M^2}{\pi^2} \sin^2 \psi
\left(1+(b^2-1) \cos^2 \psi     \right) } \right)^{1/2}
\ea
where $\psi$ is given by the minimization equation (\ref{min2}).

In the limit that $b^2-1\rightarrow 0$,  the stable solutions are given by
\ba
\label{min3}
\psi=\sqrt{1-1/q^2} \, \frac{p\, \pi}{M} 
\ea
and the energy of the system is
\ba
\label{Eng}
T_{(p,q)} =   \frac{h^2}{\lambda} 
\sqrt{  \frac{q^2}{g_s^2} + b^2\,   \left(\frac{ M}{\pi \sqrt{1-1/q^2}} \right)^2\, 
\sin^2 \left( \frac{\pi \sqrt{1-1/q^2}}{ M}\, p \right) }
\ea

Now we can compare the above energy spectrum of non-commutative $(p,q)$ strings
with the result of the dual method of wrapped D3-brane in \cite{Firouzjahi:2006vp}.  In this method the bound states of (p,q) strings are interpreted as a D3-brane with world volume electric and magnetic fluxes wrapping some two cycles inside the three sphere at the bottom of the KS throat.
The flux configurations on a D3 wrapping a 2-cycle that can induce the charge of $p$ F-strings and $q$ D-strings are
\begin{align}
\tilde{F}^{0 \mu} & = -\frac{p}{4\pi} & F_{\theta \phi} & = \frac{q}{2} 
\end{align}
where $\tilde{F}$ is the conjugate field associated with the electric field given by
${\cal{ \tilde F} }^{\mu\nu}  = - {\delta {\cal{L}} }/{\delta F_{\mu\nu}}$.

The D3-brane action is 
\ba
\label{action}
S_{D3}= -\frac{\mu_3}{g_s}\, \int d^4 \xi\sqrt{-| g_{a b} +{\cal{F}}_{ab} |} + 
\mu_3 \int \, C^{(2)} \wedge {\cal{F}} \, ,
\ea
where ${\cal{F}}_{ab}= B_{ab} + \lambda F_{ab}$. 

After integrating out fluxes and the angular components of the metric, the  Hamiltonian takes the form
\ba
\label{Hamab}
{\cal{H}} = \frac{h^2}{\lambda} \sqrt{ \frac{q^2}{g_s^2} + \frac{b^2M^2}{\pi^2} \sin^4 \psi+\left[p- \frac{M}{\pi}
\left(\psi -\frac{\sin2\psi}{2}\right)  \right]^2}\, .
\ea
with the energy
\ba
\label{eng}
E_{(p,q)} =   \frac{h^2}{\lambda} 
\sqrt{  \frac{q^2}{g_s^2} + b^2\,   \left(\frac{ M}{\pi } \right)^2\, 
\sin^2 \left( \frac{\pi }{ M}\, p \right) }
\ea

Comparing the energy in Eq.(\ref{eng}) with the energy obtained from non-commutative method
in Eq.(\ref{Eng}) we see the agreement is very close, except with the extra coefficient $\sqrt{1-1/q^2}$. 
In the limit
where $q \rightarrow \infty$ these two results agree exactly. The interpretation is that in this limit the non-commutative behavior of $\Phi^i$ becomes unimportant
and one reaches the Abelian limit.
This also happens in the examples studied in \cite{Myers:1999ps}, \cite{Myers:2003bw}. For example
when $N$ D0-branes are on top of each other in a constant four field strength $F_{tijk}$, they will
blow up into a fuzzy D2-brane. Up to the extra factor $\sqrt{1-1/N^2}$, the energy of the dielectric system matches with the energy
calculated from the dual picture of a D2-brane with magnetic fluxes wrapping a two cycle.

In the limit that $M \rightarrow \infty$, our formula Eq.(\ref{Eng}) reproduces the spectrum of  of $(p,q)$ strings in flat space-time. This is expected, since in this limit the size of $S^3$ is very large and the metric (\ref{metric0}) approximates the metric of a flat space-time. If $p=0$, we obtain the tension of $q$ D-strings with
\ba
T_{D1} =  \frac{h^2}{2 \pi \ap}\frac{q}{g_s}\, .
\ea
 
 To obtain the energy of $p$ fundamental strings one can not simply set $q=0$, since in the non-commutative method the fundamental strings were viewed as electric flux on the world volume of
 $q$ D-strings. To obtain the energy of fundamental strings one can use the S-dual picture, which
 is the subject of next section.
 

\section{The Dielectric (p,q) Strings in the S-Dual Picture }
The  {\bf{IIB} } string theory enjoys an exact symmetry, called the S-duality.
Under S-duality transformation the strongly coupled string
theory is mapped into the weakly  coupled theory. To be more specific let us denote the quantities in the S-dual picture by the superscript S. Under the S-duality transformation  $g_s  \rightarrow g_s^S=1/g_s $.
Furthermore, the $(p,q)$ system is mapped into 
$(p^S, q^S)= (-q, p)$, while $B^{(2)\, S}=-C^{(2)}, C^{(2)\, S}=B^{(2)}$ and $C^{(4)}$ remains invariant.

To obtain the energy of $p$ fundamental strings in the KS solution, we can perform an S-duality
transformation. The fundamental strings in original KS background are now $p$ coincident D-strings
in the S-dual picture. Furthermore, in the S-dual picture with $p$ coincident D-strings we can turn on
$-q$ units of electric field on their world volume to obtain the spectrum of $(-q,p)$ strings. 
Our dielectric brane method used in the previous section can now be applied again to obtain
the energy spectrum. 
Since the S-duality is exact
these two energy spectra should be equal to each other, i.e.
\ba
\label{SdualE}
T_{(p,q)}= T^S_{(-q,p)}
\ea
where $T_{(p,q)}$ is the energy of $(p,q)$ strings calculated in original KS background given in 
Eq. (\ref{Eng}) while $T^S_{(-q,p)}$ is the energy of $(p^S,q^S)= (-q, p)$ strings in the S-dual picture.

In the S-dual picture, $C^{(2)\, S}=B^{(2)}=0$ at the tip of the throat, while the Kalb-Ramond
field $B^{(2)\, S}= -C^{(2)}$ is not zero and is given by Eq. (\ref{C2}). Also $g_{MN}^S= g_s^{-1/2}\, 
g_{MN}$, so $h\rightarrow g_s^{-1/2} h$ and $R\rightarrow g_s^{-1/2} R$. There is no contribution from
$S_{CS}$ in the action, but there are new contributions in $S_{DBI}$ due to extra terms in ${Q^i_j}^S$.
We have
\ba
 {Q^{i \, S}_j}&=&  \delta^i_j + i \lambda \, [\Phi^i, \Phi^j] \, g_{k j} - i \lambda \, [\Phi^i, \Phi^j] \,  C^{(2)}_{k j}\nonumber\\
 &\equiv& Q^i_j+ \delta\, Q^i_j
\ea
where the first term, $Q^i_j$, is the same as before given in Eq. (\ref{Q}), while the last term is the change in $Q^i_j$ in the S-dual picture. One can show that
\ba
\left( \delta Q^1_1\, ,  \delta Q^2_2\, , \delta Q^3_3   \right)= -\frac{2 \lambda^2 \hat R}{R^3} 
(\psi-\frac{1}{2} \sin 2\psi)\, (\Phi_2^2+\Phi_3^2\, ,   \Phi_1^2+\Phi_3^2 \,,  \Phi_1^2+\Phi_2^2 )
\ea
and
\ba
\delta Q^i_j=   \frac{2 \lambda^2 \hat R}{R^3} \, (\psi-\frac{1}{2} \sin 2\psi)\, \Phi_j \, \Phi_i \, .
\quad \quad i\neq j
\ea
This gives
\ba
\label{detQ3}
\STr \sqrt{\det(Q^i_j)} = p \, \sqrt{ \frac{4 R^4}{\lambda^2 C_p} +\left(1 -\frac{M}{\pi \sqrt{C_p}}  (\psi-\frac{1}{2} \sin 2\psi)\right )^2 }
\ea
where $C_p=p^2-1$ is the Casimir associated with the p-dimensional irreducible representation of 
$SU(2)$.

Our action in the S-dual picture now takes the desired form Eq.(\ref{def1}) with $\dd$ given by equations
(\ref{def2}) and (\ref{detQ3}), while $\oo=0$ since $C^{(2) \, S}=0$. Furthermore, we have $-q$ units of electric fluxes so in going to Hamiltonian we set 
$D=-q$. The Hamiltonian of the system is
\ba
\label{HamS}
{\cal{H}}^S=\frac{h^2}{\lambda} \left({ \frac{q^2}{g_s^2} + \frac{1}{1-1/p^2} \frac{b^2M^2}{\pi^2} \sin^4 \psi
+\left[p- \frac{1}{\sqrt{1-1/p^2}}\frac{M}{\pi}
\left(\psi -\frac{\sin2\psi}{2}\right)  \right]^2} \right)^{1/2}\, .
\ea

This Hamiltonian has exactly the same form as the Hamiltonian (\ref{Ham2})
in the original KS background, except $C_q$ is
replaced by $C_p$, i.e. ${1}/{\sqrt{1-1/q^2}} \rightarrow  {1}/{\sqrt{1-1/p^2}}$.
Following the same steps which lead to Eq.(\ref{Eng}), for the energy of $(-q,p)$ system in the S-dual picture
in the limit $b^2-1\rightarrow 0$ we obtain
\ba
\label{Eng2}
T^S_{(q,p)} =   \frac{h^2}{\lambda} 
\sqrt{  \frac{q^2}{g_s^2} + b^2\,   \left(\frac{ M}{\pi \sqrt{1-1/p^2}} \right)^2\, 
\sin^2 \left( \frac{\pi \sqrt{1-1/p^2}}{ M}\, p \right) } \, .
\ea

Curiously enough, despite the interesting functional similarity between Eq.(\ref{Eng}) and 
Eq.(\ref{Eng2}), $T_{(p,q)} $ is not exactly equal to $T^S_{(q,p)}$, whereas under S-duality they
must be identical.
The origin of this discrepancy is the fact that under S-duality the dimension of the irreducible representation of $SU(2)$
has changed from $q$ into $p$. This has the effect of changing $C_q$ into $C_p$.
The spectrum is S-dual invariant only if $C_q$ and $C_p$ is replaced by $q^2$ and $p^2$, respectively.

One may wonder if  S-duality invariance is in fact satisfied in the dual wrapped D3-brane
method. Following the calculations of \cite{Firouzjahi:2006vp} in the S-dual picture, one can show that
indeed the energy spectrum is S-dual invariant, i.e. 
$E^S(p^S,q^S)=E(p,q)$ where $(p^S,q^S)=(-q,p)$ and
 $E(p,q)$ is given in Eq.(\ref{eng}). This is expected, since
the D3-brane is known to be manifestly S-dual invariant  \cite{Tseytlin:1996it}.

We are now able to calculate the energy of $p$ fundamental strings in KS solution.
It is given by setting $q=0$ in Eq.(\ref{Eng2}), 
which agrees with the result of \cite{Herzog:2001fq}, again except with the extra factor 
${1}/{\sqrt{1-1/p^2}}$.
The stable solutions of $p$ fundamental strings are given by
\ba
\label{min4}
\psi=\sqrt{1-1/p^2} \,\, \frac{p\, \pi}{M} \, .
\ea

The interesting new feature is that due to the extra factor $\sqrt{1-1/p^2}$ the tension of $p$ coincident 
fundamental strings does not vanish
when $p=M$. In the analysis of  \cite{Herzog:2001fq}, when $p=M$ the energy vanishes. 
From the gauge theory point of view it was argued that fundamental strings corresponds to QCD flux tubes
connecting $p$ quarks and $p$ anti-quarks of $SU(M)$ gauge theory at the tip of the conifold. When
$p=M$, these quarks and anti-quarks confine to create the baryon and anti-baryon and the fluxes disappear.
If one to take this argument seriously in the light of AdS/CFT correspondence, then this requires that
the square root in Eq.(\ref{min4}) should be replaced by unity which is equivalent to replacing $C_p$ by $p^2$. This is also supported by the S-duality
invariance demand as described above.

\section{Discussion}
Using the dielectric brane prescription of Myers the $(p,q)$ strings spectrum is calculated in KS throat.
With $q$ D-strings on top of each other, they blow up into a fuzzy two sphere. This represents a
bound state of  a D3-brane with $q$ D1-branes smeared on it. The F-strings are
incorporated in the model by turning on electric fluxes on the world volume of D-strings.
The result for energy spectrum is in interesting agreement with the dual picture, where the $(p,q)$ string is interpreted
as a wrapped D3-brane with $p$ units of electric flux and $q$ units of magnetic flux in its world volume.
The agreement is exact when $q \rightarrow \infty$, as expected. 

Since S-duality is an exact symmetry of {\bf IIB} string theory, one expects that the energy spectrum calculated in the
S-dual picture is the same as in the original picture. Although the energy spectrum has exactly the same
functional form in both pictures, however there are interesting parametric discrepancies. This is due
to an interchange of the Casimir in two pictures. In the original picture $\Phi^i$ represent a $q$-dimensional irreducible representation of $SU(2)$ algebra, whereas in the S-dual picture they belong to a 
$p$-dimensional irreducible representation of $SU(2)$ algebra. The agreement is exact only when
both $p$, $q \rightarrow \infty$. 

This raises the question wether the dielectric brane prescription of   \cite{Myers:1999ps} is invariant under
S-duality in {\bf IIB} string theory. Obviously, to recover the S-duality in this calculations one needs to
``effectively"  replace $C_p$ and $C_q$ by $p^2$ and $q^2$, respectively. 
One may try to add some new terms in equations  (\ref{SDBI}) and (\ref{SCS}) to restore the S-duality.
Although this is a possibility, but it should be performed in way not to spoil the T-duality invariance of the
original prescription. The most probable cause of S-duality violation may be the symmetric trace prescription used in (\ref{SDBI}) and (\ref{SCS}). It is known that the symmetric trace prescription
does not agree with the full effective string action \cite{Hashimoto:1997gm}, \cite{Bain:1999hu} and corrections of order $F^6$ and higher in the world-volume field strength should be added to the action.
It is interesting to know wether a better prescription for the trace operation in the non-Abelian action can
resolve the problem. 
 If so, the benefits are two folds: not only the S-duality is restored, but also the dielectirc
calculation matches exactly  with the result of dual method for any value of $q$. This happens in
$(p,q)$ string case here and also for the case in \cite{Myers:1999ps} where $N$ D0-branes blow up into a D2 brane.

\begin{acknowledgments}{I thank  Neil Barnaby, 
Keshav Dasgupta, Igor Klebanov, Louis Leblond, Rob Myers,
Omid Saremi and Henry Tye for useful discussions and comments. This work is supported in part by  NSF under Grant No. PHY-009831 and 
NSERC under Grant No. 204540. }\\
\end{acknowledgments}


\section*{References}
\begin{thebibliography}{}

\bibitem{Spergel:2006hy}
  D.~N.~Spergel {\it et al.},
  ``Wilkinson Microwave Anisotropy Probe (WMAP) three year results:
  Implications for cosmology,'' astro-ph/0603449.
  
\bibitem{Dasgupta:1999ss}
  K.~Dasgupta, G.~Rajesh and S.~Sethi,
  JHEP {\bf 9908}, 023 (1999)
  [arXiv:hep-th/9908088].

\bibitem{Giddings:2001yu}
S.~B.~Giddings, S.~Kachru and J.~Polchinski,
``Hierarchies from fluxes in string compactifications,''
Phys.\ Rev.\ D {\bf 66}, 106006 (2002), hep-th/0105097.

\bibitem{Kachru:2003aw}
S.~Kachru, R.~Kallosh, A.~Linde and S.~P.~Trivedi,
'' De Sitter vacua in string theory'',
Phys.\ Rev.\ D {\bf 68}, 046005 (2003),
hep-th/0301240.

\bibitem{Randall:1999ee} L.~Randall and R.~Sundrum,
  " A large mass hierarchy from a small extra dimension", 
  Phys.\ Rev.\ Lett.\  {\bf 83}, 3370 (1999),
 hep-ph/9905221.
 
\bibitem{Klebanov:2000hb}
I.~R.~Klebanov and M.~J.~Strassler,
``Supergravity and a confining gauge theory: Duality cascades and
chiSB-resolution of naked singularities,''
JHEP {\bf 0008}, 052 (2000), hep-th/0007191.

\bibitem{dvali-tye} G.~Dvali and S.-H.H.~Tye,
"Brane Inflation",
Phys. Lett. {\bf B450} (1999) 72, hep-ph/9812483.

\bibitem{collection}
C.~P.~Burgess, M.~Majumdar, D.~Nolte, F.~Quevedo, G.~Rajesh 
and R.~J.~Zhang, JHEP {\bf 07} (2001) 047, hep-th/0105204\, ; 
G.~R.~Dvali, Q.~Shafi and S.~Solganik,
``D-brane inflation,''
hep-th/0105203.    

\bibitem{Kachru:2003sx}
S.~Kachru, R.~Kallosh, A.~Linde, J.~Maldacena, L.~McAllister
and S.~P.~Trivedi, 
'' Towards inflation in string theory'',
JCAP {\bf 0310} (2003) 013, hep-th/0308055.

\bibitem{Firouzjahi:2003zy}
H.~Firouzjahi and S.-H.~H.~Tye,
``Closer towards inflation in string theory,''
Phys.\ Lett.\ B {\bf 584}, 147 (2004), hep-th/0312020 \, ;
C.~P.~Burgess, J.~M.~Cline, H.~Stoica and F.~Quevedo,
``Inflation in realistic D-brane models,''
JHEP {\bf 0409}, 033 (2004), hep-th/0403119\, ;   A.~Buchel and R.~Roiban, ``Inflation in warped geometries,''
  Phys.\ Lett.\ B {\bf 590}, 284 (2004)
  [arXiv:hep-th/0311154]\, ;
  N.~Iizuka and S.~P.~Trivedi,
  ``An inflationary model in string theory,''
  Phys.\ Rev.\ D {\bf 70}, 043519 (2004), [arXiv:hep-th/0403203].
  
   \bibitem{Sarangi:2002yt}
S.~Sarangi and S.-H.~H.~Tye,
``Cosmic string production towards the end of brane inflation,''
Phys.\ Lett.\ B {\bf 536}, 185 (2002), hep-th/0204074.

\bibitem{Jones:2003da}
N.~T.~Jones, H.~Stoica and S.-H.~H.~Tye,
'' The production, spectrum and evolution of cosmic strings 
in brane inflation'',
Phys.\ Lett.\ B {\bf 563}, 6 (2003), hep-th/0303269.

\bibitem{Copeland:2003bj}
E.~J.~Copeland, R.~C.~Myers and J.~Polchinski,
``Cosmic F- and D-strings,''
JHEP {\bf 0406}, 013 (2004), hep-th/0312067 \, ;
L.~Leblond and S.-H.~H.~Tye,
``Stability of D1-strings inside a D3-brane,''
JHEP {\bf 0403}, 055 (2004), hep-th/0402072.

\bibitem{Firouzjahi:2005dh}
  H.~Firouzjahi and S.-H.~H.~Tye,
" Brane inflation and cosmic string tension in superstring theory,"
 JCAP {\bf 0503}, 009 (2005),
 hep-th/0501099.
 
\bibitem{Shandera:2006ax}
  S.~E.~Shandera and S.~H.~Tye, ``Observing brane inflation,''
  JCAP {\bf 0605}, 007 (2006)
  [arXiv:hep-th/0601099].

\bibitem{Wyman:2005tu}
  M.~Wyman, L.~Pogosian and I.~Wasserman, ``Bounds on cosmic strings from WMAP and SDSS,''
  Phys.\ Rev.\ D {\bf 72}, 023513 (2005)
  [Erratum-ibid.\ D {\bf 73}, 089905 (2006)]
  [arXiv:astro-ph/0503364].
  
\bibitem{Seljak:2006hi}
  U.~Seljak and A.~Slosar, ``B polarization of cosmic microwave background as a tracer of strings,''
  arXiv:astro-ph/0604143.

\bibitem{Firouzjahi:2006vp}
  H.~Firouzjahi, L.~Leblond and S.~H.~Henry Tye,
  ``The (p,q) string tension in a warped deformed conifold,''
  JHEP {\bf 0605}, 047 (2006)
  [arXiv:hep-th/0603161].

 \bibitem{Myers:1999ps}
  R.~C.~Myers,
  ``Dielectric-branes,''
  JHEP {\bf 9912}, 022 (1999)
  [arXiv:hep-th/9910053].
  
\bibitem{Myers:2003bw}
  R.~C.~Myers,
  ``Nonabelian phenomena on D-branes,''
  Class.\ Quant.\ Grav.\  {\bf 20}, S347 (2003)
  [arXiv:hep-th/0303072].


\bibitem{Candelas:1989js}
P.~Candelas and X.~C.~de la Ossa,
"Comments On Conifolds,"
Nucl.\ Phys.\ B {\bf 342}, 246 (1990).
    
\bibitem{Minasian:1999tt}
R.~Minasian and D.~Tsimpis,
"On the geometry of non-trivially embedded branes,"
Nucl.\ Phys.\ B {\bf 572}, 499 (2000), hep-th/9911042.

\bibitem{Ohta:1999we}
K.~Ohta and T.~Yokono,
" Deformation of conifold and intersecting branes,"
JHEP {\bf 0002}, 023 (2000), hep-th/9912266.

\bibitem{Herzog:2001xk}
C.~P.~Herzog, I.~R.~Klebanov and P.~Ouyang,
" Remarks on the warped deformed conifold,"
hep-th/0108101.
  
  \bibitem{Herzog:2001fq}
  C.~P.~Herzog and I.~R.~Klebanov,
  ``On string tensions in supersymmetric SU(M) gauge theory,''
  Phys.\ Lett.\ B {\bf 526}, 388 (2002)
  [arXiv:hep-th/0111078].
  
    
  \bibitem{Thomas:2006ud}
  S.~Thomas and J.~Ward,
  ``Non-Abelian (p,q) strings in the warped deformed conifold,''
  arXiv:hep-th/0605099.
  
\bibitem{Tseytlin:1997cs}
  A.~A.~Tseytlin,
  Nucl.\ Phys.\ B {\bf 501}, 41 (1997)
  [arXiv:hep-th/9701125].
  
  \bibitem{Constable:1999ac}
  N.~R.~Constable, R.~C.~Myers and O.~Tafjord,
  ``The noncommutative bion core,''
  Phys.\ Rev.\ D {\bf 61}, 106009 (2000)
  [arXiv:hep-th/9911136].
  
  
\bibitem{Tseytlin:1996it}
  A.~A.~Tseytlin,
  Nucl.\ Phys.\ B {\bf 469}, 51 (1996)
  [arXiv:hep-th/9602064]; \, 
  M.~B.~Green and M.~Gutperle,
  Phys.\ Lett.\ B {\bf 377}, 28 (1996)
  [arXiv:hep-th/9602077].
  
\bibitem{Hashimoto:1997gm}
  A.~Hashimoto and W.~I.~Taylor,
  Nucl.\ Phys.\ B {\bf 503}, 193 (1997)
  [arXiv:hep-th/9703217].
  
\bibitem{Bain:1999hu}
  P.~Bain,
  arXiv:hep-th/9909154.
  
  
  
  
  
  


\end{thebibliography}
\end{document}